\documentstyle[aps]{revtex}  
\begin{document} 
\twocolumn 
\wideabs{
\title{ 
Zeta functions, renormalization group equations, and
the effective action
} 
\author{ 
David Hochberg$^{+}$, Carmen Molina--Par\'{\i}s$^{++}$,  
Juan P\'erez--Mercader$^{+}$, and Matt Visser$^{+++}$ 
} 
\address{ 
$^{+}$Laboratorio de Astrof\'{\i}sica Espacial y F\'{\i}sica 
Fundamental, Apartado 50727, 28080 Madrid, Spain\\ 
$^{++}$Theoretical Division, Los Alamos National Laboratory, 
Los Alamos, New Mexico 87545, USA\\ 
$^{+++}$Physics Department, Washington University, 
Saint Louis, Missouri 63130-4899, USA\\ 
} 
\date{30 September 1998; Printed \today}
\maketitle 

{\small {\bf Abstract:} We demonstrate how to extract all the one-loop
renormalization group equations for {\em arbitrary} quantum field
theories from knowledge of an appropriate Seeley--DeWitt
coefficient. By formally solving the renormalization group equations
to one loop, we renormalization group {\em improve} the classical
action, and use this to derive the leading-logarithms in the one-loop
effective action for {\em arbitrary} quantum field theories.}
\pacs{xx.yy.zz}
} 
\newcommand{\Str}{\mathop{\mathrm{Str}}} 
\newcommand{\tr}{\mathop{\mathrm{tr}}} 
\newcommand{\define}{\mathop{\stackrel{\rm def}{=}}}
\def\d{{\mathrm{d}}}
\def\a2{a_{d/2}}
\def\implies{\Rightarrow}
\def\dirac{\gamma^\mu (\partial_\mu - A_\mu)}
\def\half{ {\scriptstyle{1\over2}} }

{\em Introduction:} It is well-known that {\em any} quantum field
theory (QFT) can be renormalized to one loop via the magic of zeta
function techniques
\cite{Dowker,Hawking,Gibbons,DeWitt,Hawking2,Centenary,BVW}.  These
techniques are so powerful that they hide all of the divergence
structure in the woodwork, which has caused the community to miss a
very important point: The one-loop renormalization group equations
(RGE's) for {\em arbitrary} QFT's can be
extracted from knowledge of an appropriate Seeley--DeWitt coefficient
({\em aka} Hamidew coefficient, Schwinger--DeWitt coefficient,
Seeley--Gilkey coefficient, Hadamard coefficient, Heat Kernel
coefficient). Assuming only that the kinetic energy terms are
quadratic in derivatives, and that spacetime has $d$ dimensions, the
appropriate Seeley--DeWitt coefficient is $a_{d/2}$, a quantity that
is already known to be of paramount importance in that it governs the
conformal scale anomaly~\cite{Dowker,Centenary,BVW}. For the purposes
of this Letter the central observation is that $a_{d/2}$ also governs
the one-loop logarithmic divergences, and so controls the running of
the coupling constants at one-loop order. By expanding the classical
action and Seeley--DeWitt coefficients in terms of primitive symmetry
invariants, the one-loop RGE's can easily (if formally)
be written down for arbitrary QFT's. It is then easy to see that {\em
all} one-loop beta functions vanish in odd-dimensional spacetimes, so
that there is no running of the coupling constants at one-loop
order. For even-dimensional spacetimes, with an appropriate set of
conventions, all beta functions can be written as homogeneous
multinomials of order $d/2$.

By formally solving the RG equations, we can one-loop {\em improve}
the classical action. The improved classical action then allows us to
extract information about the one loop effective action without ever
resorting to explicit Feynman diagram calculations. In particular, the
form of the leading-logarithmic contributions to the effective action
is completely specified in terms of the $\a2$ Seeley--DeWitt
coefficient.  These coefficients are tabulated in many places, and
computation is now essentially automated~\cite{Encyclopedia}, making
this observation of considerable practical importance.


{\em Effective action and RGE at one loop:} Consider an arbitrary
quantum field $\phi(\vec x,t)$ governed by a classical action
$S[\phi,\lambda_i]$. The field $\phi$ may be scalar, spinor, or (with
suitable caveats) even a gauge field.  The set $\{\lambda_i\}$ denotes
the complete set of (generalized) coupling constants in the theory.
It is a standard result that (in terms of bare quantities) the
one-loop effective action is
\begin{eqnarray}
\Gamma[\phi;\phi_0] &=& 
S[\phi,\lambda_i] - S[\phi_0,\lambda_i] 
\nonumber\\
&+& {\hbar\over2} \Str
\left\{\ln\det {S_2[\phi,\lambda_i]\over\mu^2} 
- \ln\det {S_2[\phi_0,\lambda_i]\over\mu^2} \right\}
\nonumber\\
&+&O(\hbar^2).
\end{eqnarray}
Here $\Str$ denotes a ``supertrace'', a sum over all bose and fermi
fields in the theory with a $+$ sign for bose fields and a $-$ for
fermi fields.  Spin degeneracy factors are subsumed into the
determinant. $S_2[\phi]$ denotes $\delta^2 S[\phi] /
(\delta\phi(x)\,\delta\phi(y))$, which is a second-order differential
operator that governs the Gaussian fluctuations. (For a unified
notation, fermion determinants can always be converted to second order
by squaring before taking the determinant. Also, for gauge fields one
should be careful to include terms due to gauge-breaking, and the
unitarity preserving ghosts~\cite{Savvidy,Avramidi}.) The arbitrary
parameter $\mu$ has been introduced for purely dimensional reasons (to
keep the argument of the logarithm dimensionless). Furthermore $S_2$
is second-order because in this Letter we are interested in QFT's; for
the considerably more complicated non-quantum field theories
associated with stochastic differential equations~\cite{Zinn-Justin}
this particular assumption must be
modified~\cite{HMPV-in-preparation}.  Finally $\phi_0$ is some
suitable background field (a classical solution of the equations of
motion for zero source), typically a minimum of the bare potential, a
zero gauge field strength, Minkowski spacetime, or even Schwarzschild
spacetime. The above is of course a divergent quantity which has to be
regularized and renormalized. Invoking completely standard machinery,
to be found in many QFT textbooks~\cite{Zinn-Justin,Weinberg,Collins},
we do so with the result that (now in terms of renormalized
quantities)
\begin{eqnarray}
&&\Gamma[\phi;\phi_0] = 
S[\phi,\lambda_i(\mu)] - S[\phi_0,\lambda_i(\mu)] 
\nonumber\\
&& \qquad + {\hbar\over2} 
\Str \left\{\ln\det {S_2[\phi,\lambda_i(\mu)]\over\mu^2} 
- \ln\det {S_2[\phi_0,\lambda_i(\mu)]\over\mu^2} \right\}
\nonumber\\
&&\qquad +O(\hbar^2).
\end{eqnarray}
The coupling constants now in general ``run'' with the renormalization
scale $\mu$. (The only slightly non-standard thing we have done here
is to avoid using the wavefunction renormalization $Z(\mu)$ to rescale
the quantum field; instead we find it more convenient to view
wavefunction renormalization as just another coupling constant.) The
{\em exact} renormalization group equations are simply the statement
that the effective action does not depend on the renormalization
scale, {\em i.e.},
\begin{equation}
\label{RGE:0}
{\mu\;\d\Gamma[\phi;\phi_0]\over\d\mu} = 0.
\end{equation}
Now from zeta function technology (or with a little more work from any
other regularization and renormalization scheme), and assuming for
simplicity the lack of infra-red divergences, we have the {\em exact}
mathematical result that
\cite{Dowker,Hawking,Gibbons,DeWitt,Hawking2,Centenary,BVW}
\begin{eqnarray}
&&\ln\det {S_2[\phi,\lambda_i(\mu)]\over\mu^2} =
\ln\det {S_2[\phi,\lambda_i(\mu)]\over\mu_0^2} 
\nonumber\\
&& \qquad + 
{1\over(4\pi)^{d/2}} \int \d^d x \; a_{d/2}[\phi,\lambda_i(\mu)] \; 
\ln\left({\mu\over\mu_0}\right).
\end{eqnarray}
To be even more explicit, this entails
\begin{eqnarray}
&&\ln\det {S_2[\phi,\lambda_i(\mu)]\over\mu^2} =
\ln\det {S_2[\phi,\lambda_i(\mu_0)]\over\mu_0^2} 
\nonumber\\
&& \quad + 
{1\over(4\pi)^{d/2}} \int \d^d x \; a_{d/2}[\phi,\lambda_i(\mu)] \;
\ln\left({\mu\over\mu_0}\right)
+O(\hbar).
\end{eqnarray}
So, inserting this into the exact RGE we obtain to one-loop order
\begin{eqnarray}
\label{RGE:1}
&&{\mu\;\d S[\phi,\lambda_i(\mu)]\over\d\mu} - 
{\mu\;\d S[\phi_0,\lambda_i(\mu)]\over\d\mu} = 
\nonumber\\
&& \quad - 
{\hbar\over2(4\pi)^{d/2}} \int \d^d x  \Str
\left\{ a_{d/2}[\phi,\lambda_i(\mu)] - a_{d/2}[\phi_0,\lambda_i(\mu)] \right\}
\nonumber\\
&& \quad + O(\hbar^2).
\end{eqnarray}
Equivalently
\begin{eqnarray}
\label{RGE:2}
&&\left\{ 
{\d S[\phi,\lambda_i(\mu)]\over\d\lambda_i(\mu)} -
{\d S[\phi_0,\lambda_i(\mu)]\over\d\lambda_i(\mu)} \right\} 
{\mu\;\d\lambda_i(\mu)\over\d\mu} = 
\nonumber\\
&& \quad -
{\hbar\over2(4\pi)^{d/2}} \int \d^d x \Str
\left\{ a_{d/2}[\phi,\lambda_i(\mu)] - a_{d/2}[\phi_0,\lambda_i(\mu)] \right\}
\nonumber\\
&& \quad + O(\hbar^2).
\end{eqnarray}
Extracting beta-functions from this is now completely straightforward:
We pick off terms of the same functional form from both sides of the
above~\cite{Fujimoto,Gato,HMPV-rgi}. Results may be simplified
drastically by choosing an appropriate set of conventions. Let
$\Phi_i$ be a basis of elementary terms in the classical action
constrained only by symmetry. For instance for a scalar theory we
would typically have $\Phi_0 = \half (\partial\phi)^2$ and $\Phi_n =
{\scriptstyle{1\over n!}}\phi^n$, for fermions we would take $\Phi_0 =
\bar\psi[\dirac]\psi$, $\Phi_2=m\;\bar\psi\psi$, and for gauge
theories $\Phi_0=F^2$ and $\Phi_1=F\tilde F$. For mixed theories we
just have to rearrange the indices, and without loss of generality we
can adopt the convention that the action is linear in this basis
and in the generalized coupling constants:
\begin{equation}
S[\phi,\lambda_i(\mu)] = \sum_i \lambda_i(\mu) \int \d^d x \; \Phi_i.
\end{equation}
This is only a convention, it is not a restriction on the class of
theories considered. Given this choice of basis we can also
expand the Seeley--DeWitt coefficient as
\begin{equation}
\Str\left( a_{d/2}[\phi,\lambda_i(\mu)] \right) = 
\sum_i \kappa_i(\lambda_j(\mu)) \; \Phi_i.
\end{equation}
That the same set of elementary terms can be used to expand both the
classical action {\em and} the integrated Seeley--DeWitt coefficient
is a consequence of renormalizability. Specifically, for
renormalizable and super-renormalizable theories the counterterms are
by definition equal to or fewer than the elementary terms in the
classical action, which implies that the Seeley--DeWitt coefficient is
expandable using the elementary terms occurring in the classical action
as a basis. For non-renormalizable theories this fails, since there
are terms in the integrated Seeley--DeWitt coefficient that do not
appear in the classical action.  (The Seeley--DeWitt coefficient will
often contain total derivatives, such as $\nabla^2\phi$, which could
be added to the classical Lagrangian without affecting the classical
action, and so can be omitted altogether since we are only really
interested in spacetime integrals.) With all these conventions in
place, our (slightly nonstandard) one-loop beta functions are
\begin{equation}
\beta_i(\lambda_j) 
\define {\mu\;\d\lambda_i(\mu)\over\d\mu} 
= - {\hbar\over2(4\pi)^{d/2}} \; \kappa_i(\lambda_j(\mu)) + O(\hbar^2).
\end{equation}
An immediate consequence is that {\em all\,} of our beta functions
vanish to one loop in odd-dimensional spacetimes, simply because the
Seeley--DeWitt coefficient vanishes in odd-dimensional
spacetimes~\cite{BVW}. (This is intimately related to the vanishing of
the conformal anomaly in odd-dimensional
spacetimes~\cite{Birrell-Davies}.) This statement is not limited to
flat space and continues to hold true even for QFT's defined on curved
spacetimes. It will however fail in general for manifolds with
boundary. For manifolds with boundary the classical action must
contain both bulk and surface contributions, and while the coupling
constants associated with the bulk action do not run at one loop,
those coupling constants appearing in the surface action will
generally run at one loop. We are {\em not} asserting that all
odd-dimensional (spacetime) theories are one-loop finite, but the much
more modest claim that all odd-dimensional (spacetime) theories are
one-loop non-running.

Explicit calculation will quickly verify that all one-loop beta
functions vanish for model theories such as QED${}_3$ and
$\lambda(\phi^4)_3$. In contrast, within the context of the
$\epsilon$--expansion, one-loop beta functions for
QED${}_{4-\epsilon}$ and $\lambda(\phi^4)_{4-\epsilon}$ do not vanish,
but these beta functions should not be trusted for $\epsilon=1$.

The attentive reader might profitably wonder where we have hidden all
the anomalous dimensions? All anomalous dimensions have been converted
into beta functions via the schema
\begin{equation}
\gamma_Z \define {\mu\;\d\ln Z(\mu)\over\d\mu} 
\quad \implies \quad
\beta_Z \define {\mu\;\d Z(\mu)\over\d\mu} = Z \;\gamma_Z.
\end{equation}
The inverse transformation is trivial, but we prefer a schema that
handles everything in a unified fashion.

Before we turn to issues of improving the classical action and
deriving the leading-logarithm contributions to the effective action,
is there anything more we can say about the $\kappa_i(\lambda_j)$
without resorting to explicit calculations? Start by observing that the
Jacobi field operator is by definition linear in the couplings
\begin{equation}
S_2[\phi,\lambda_i(\mu)] = \sum_i \lambda_i(\mu) 
\left.{\delta^2\Phi_i\over\delta\phi(x)\delta\phi(y)}\right|_{\phi},
\end{equation}
and note that the $d/2$'th Seeley--DeWitt coefficient is homogeneous
in the Jacobi field operator
\begin{equation}
\a2(\alpha S_2) = \a2(S_2).
\end{equation}
This can be derived from the definition of $\a2$ in terms of the
short-time expansion of the heat kernel, and implies
that for {\em all\,} QFT's at one-loop
\begin{equation}
\beta_i(\alpha \lambda_j) = \beta_i(\lambda_j).
\end{equation}
This homogeneity property is often enough to completely pin down the
form (if not the coefficients) of the one-loop beta functions. For
instance for pure gauge theories (no matter fields), our conventions
imply that we must write $S(\phi,\lambda) = F^2/g^2$. Since there is
only one coupling constant present in the theory ($\lambda_0=1/g^2$)
the homogeneity result implies
\begin{equation}
{\mu \; \d(1/g^2)\over \d\mu} = \hbar\; k + O(\hbar^2).
\end{equation}
Here $k$ is now some constant independent of $g$. That is: gauge
symmetry plus the analysis of this Letter is enough to specify the
{\em form} of the one-loop beta function completely.

Scalar field theory provides another useful example: In this case the
coupling constant $\lambda_0$ attached to the kinetic energy term
$(\Phi_0)$ plays a special role, and the homogeneity relation can
always be used to scale it out of the $\a2$ coefficient. Furthermore,
there are well-known recursion relations for calculating the
Seeley--DeWitt coefficients in terms of the Jacobi field operator. The
key point is that $\a2$ contains terms of the type
$[(S_2)^{d/2}I(x,x')]$, plus lower powers of $S_2$, plus derivative
terms that integrate to zero. (Here $I(x,x')$ is the parallel
displacement operator, and the square brackets indicate the
coincidence limit $x'\to x$.)  This implies that $\int \a2 $ is a
multinomial in $(\lambda_i/\lambda_0)$ of order $d/2$ ({\em i.e.}
containing terms up to $\lambda_i^{d/2}$). More specifically, for 2,
4, and 6 dimensions, the beta functions (using our conventions) must
always be of the form
\begin{eqnarray}
\beta_i(\lambda_j;d=2) 
&=& - {\hbar\over8\pi} \kappa_i{}^j \; 
{\lambda_j\over\lambda_0} 
+ O(\hbar^2),
\\
\beta_i(\lambda_j; d=4) 
&=& - {\hbar\over32\pi^2} \kappa_i{}^{jk} \; 
{\lambda_j\over\lambda_0} \; 
{\lambda_k\over\lambda_0}
+ O(\hbar^2),
\\
\beta_i(\lambda_j;d=6) 
&=& - {\hbar\over128\pi^3} \kappa_i{}^{jkl} \; 
{\lambda_j\over\lambda_0} \; 
{\lambda_k\over\lambda_0} \; 
{\lambda_l\over\lambda_0}
+ O(\hbar^2),
\end{eqnarray}
with the obvious pattern holding for higher dimensions. This can be
checked against explicit computations for standard theories
(see {\em e.g.} Collins~\cite{Collins}, or~\cite{HMPV-rgi}) which show
that the $\kappa$'s are simple rational numbers. For $\lambda
(\phi^4)_4$ these constraints can be used to completely fix the {\em
form} of the one-loop beta functions. This structure can also be
justified via rather general Feynman diagram considerations: the
beta-functions at one loop are a reflection of the logarithmic
divergences; in $d$ dimensions we get one-loop logarithmic divergences
only from a polygonal loop with $d/2$ propagators (and so $d/2$
vertices). With our conventions each one of these vertices must
contain exactly one $\lambda_i$, and each propagator a $1/\lambda_0$,
which completes the proof.


{\em RG improvement and consistency check:} Suppose now we have
extracted the RGE's and have integrated them up to obtain
$\lambda_i(\mu) = f_i(\lambda_j(\mu_0),\mu/\mu_0)$. The improved
action is defined by inserting these running parameters into the
classical action
\begin{equation}
S_{\mathrm{improved}}(\phi,\lambda_i(\mu)) 
\define 
S(\phi,\lambda_i\to\lambda_i(\mu)).
\end{equation}
Now the RGE's [{\em cf} Eq. (\ref{RGE:2})] have been carefully arranged
so that
\begin{eqnarray}
&&
S_{\mathrm{improved}}(\phi,\lambda_i(\mu)) 
= 
S_{\mathrm{improved}}(\phi,\lambda_i(\mu_0)) 
\nonumber\\
&&\qquad  +
{\hbar\over2(4\pi)^{d/2}} \int \d^d x \; \Str a_{d/2}[\phi,\lambda_i(\mu)] \;
\ln\left({\mu\over\mu_0}\right)
\nonumber\\
&&\qquad
+O(\hbar^2).
\end{eqnarray} 
To one-loop order the effective action satisfies the
consistency condition
\begin{eqnarray}
&&\Gamma[\phi;\phi_0] 
= 
S[\phi,\lambda_i(\mu)] - S[\phi_0,\lambda_i(\mu)]
\nonumber\\ 
&& \qquad + {\hbar\over2} \Str
\left\{\ln\det {S_2[\phi,\lambda_i(\mu)]\over\mu^2} 
- \ln\det {S_2[\phi_0,\lambda_i(\mu)]\over\mu^2} \right\}
\nonumber\\
&&\qquad + O(\hbar^2),
\\
&&\qquad =
S[\phi,\lambda_i(\mu_0)] - S[\phi_0,\lambda_i(\mu_0)] 
\nonumber\\
&&\qquad + {\hbar\over2} \Str
\left\{\ln\det {S_2[\phi,\lambda_i(\mu_0)]\over\mu_0^2} 
- \ln\det {S_2[\phi_0,\lambda_i(\mu_0)]\over\mu_0^2} \right\}
\nonumber\\
&&\qquad + O(\hbar^2).
\end{eqnarray}
This verifies, as it should, that physics is independent of the choice
of renormalization scale $\mu$.

{\em Leading logarithms:} The general solution of the RGE is
\begin{eqnarray}
\label{E:ansatz}
&&\Gamma[\phi;\phi_0] 
= 
S[\phi,\lambda_i(\mu)] - S[\phi_0,\lambda_i(\mu)] 
 + {\hbar\over2d(4\pi)^{d/2}} 
\nonumber\\
&&
\qquad
\times \int \d^d x \Bigg\{ \Str \Bigg(
\a2[\phi,\lambda_i(\mu)]\ln{\a2[\phi,\lambda_i(\mu)]\over\mu^d} 
\nonumber\\
&& \qquad -
\a2[\phi_0,\lambda_i(\mu)]\ln{\a2[\phi_0,\lambda_i(\mu)]\over\mu^d} \Bigg)
\nonumber\\
&&
\qquad + X[\lambda_i(\mu),\Phi_i] \Bigg\}
+O(\hbar^2).
\end{eqnarray}
Here we use the fact that the RGE [Eq. (\ref{RGE:0}), or equivalently
Eq. (\ref{RGE:2})] is a quasi-linear first order partial differential
equation~\cite{Math} and adjust integration constants in a convenient
way.  (Some special cases are discussed in~\cite{HMPV-rgi}.) The
integration constant $X$ is constrained by the facts that (1) it {\em
cannot} depend explicitly on $\mu$, and (2) by dimensional analysis
and renormalizability, to be of the form
\begin{equation}
X[\lambda_i(\mu),\Phi_i] = 
\sum_i \epsilon_i(\lambda_j(\mu), \Phi_j) \; \lambda_i(\mu) \; \Phi_i.
\end{equation}
Here the $\epsilon_i(\lambda_j(\mu), \Phi_j)$ are dimensionless
functions of the indicated variables. This is sometimes sufficient to
specify the $\epsilon_i$ completely. For instance for scalar field
theories in the constant field limit ({\em i.e.} the effective
potential) the $\epsilon_i$ are known to be
constants~\cite{HMPV-rgi}, and so simply correspond to finite
renormalization ambiguities. Thus equation (\ref{E:ansatz}) in this
case provides the {\em exact} one-loop effective potential. This also
holds for fermion plus scalar systems with Yukawa interactions, but
once background gauge fields are switched on there are too many
dimensionfull operators present to usefully constrain
$X[\lambda_i,\Phi_i]$~\cite{HMPV-rgi}.

More generally we can appeal to a variant of the decoupling
theorem~\cite{Collins}, by noting that $\a2$ behaves like a mass
term for the Gaussian fluctuations. Thus an expansion in strong fields
is equivalently an expansion in large masses and so the decoupling
theorem justifies the result that
\begin{eqnarray}
\label{E:strong}
&&\Gamma[\phi;\phi_0] = S[\phi,\lambda_i(\mu)] - S[\phi_0,\lambda_i(\mu)]
+ {\hbar\over2d(4\pi)^{d/2}} 
\nonumber\\
&&
\qquad \times \int \d^d x  \Str \Bigg\{
\a2[\phi,\lambda_i(\mu)] 
\Bigg[ \ln{\a2[\phi,\lambda_i(\mu)]\over\mu^d}
\nonumber\\
&& \qquad \qquad +
O\left({\a2[\phi_0,\lambda_i(\mu)]\over\a2[\phi,\lambda_i(\mu)]}\right)
\Bigg] \Bigg\}
+O(\hbar^2).
\end{eqnarray}
%


{\em Discussion:} One-loop physics is important for two reasons: it is
relatively easy to calculate at one loop {\em and} it is often the
case that one-loop is sufficient to extract most of the important
physics from a problem. While it is well known that the zeta function
techniques and the Seeley--DeWitt expansion have a lot to say about
one-loop physics, the true breadth and generality of these results has
not been elucidated before now. In this Letter we have shown that
essentially all of one-loop physics for all QFT's ({\em i.e.}, systems
with fluctuations governed by a second-order differential operator)
can be extracted from the appropriate Seeley--DeWitt coefficient,
$\a2$. The analysis also puts very strong constraints on the form of
the one-loop beta functions without ever having to resort to a
specific Feynman diagram calculation.

 
\end{document}